\documentclass[conference]{IEEEtran}
\IEEEoverridecommandlockouts
\usepackage{cite}
\usepackage{amsmath,amssymb,amsfonts}
\usepackage{algorithmic}
\usepackage{graphicx}
\usepackage{textcomp}
\usepackage{xcolor}
\usepackage{multirow}
\usepackage{url}
\usepackage{enumerate}

\usepackage{balance}
\def\BibTeX{{\rm B\kern-.05em{\sc i\kern-.025em b}\kern-.08em
    T\kern-.1667em\lower.7ex\hbox{E}\kern-.125emX}}
\begin{document}

\title{Keywords Guided Method Name Generation\\
}

\author{
	\IEEEauthorblockN{Fan Ge}
	\IEEEauthorblockA{\textit{School of Computer Science and Engineering} \\
		\textit{Central South University}\\
		Changsha, China \\
		lesliege@csu.edu.cn}
	\and 
	\IEEEauthorblockN{Li Kuang* \thanks{* Li Kuang is the corresponding author.}}
	\IEEEauthorblockA{\textit{School of Computer Science and Engineering} \\
		\textit{Central South University}\\
		Changsha, China \\
		kuangli@csu.edu.cn}
}
\maketitle

\begin{abstract}
High quality method names are descriptive and readable, which are helpful for code development and maintenance. The majority of recent research suggest method names based on the text summarization approach. They take the token sequence and abstract syntax tree of the source code as input, and generate method names through a powerful neural network based model. However, the tokens composing the method name are closely related to the entity name within its method implementation. Actually, high proportions of the tokens in method name can be found in its corresponding method implementation, which makes it possible for incorporating these common shared token information to improve the performance of method naming task. Inspired by this key observation, we propose a two-stage keywords guided method name generation approach to suggest method names. Specifically,
we decompose the method naming task into two subtasks, including keywords extraction task and method name generation task. For the keywords extraction task, we apply a graph neural network based model to extract the keywords from source code. For the method name generation task, 
we utilize the extracted keywords to guide the method name generation model. We apply a dual selective gate in encoder to control the information flow, and a dual attention mechanism in decoder to combine the semantics of input code sequence and keywords. Experiment results on an open source dataset demonstrate that keywords guidance can facilitate method naming task, which enables our model to outperform the competitive state-of-the-art models by margins of 1.5\%-3.5\% in ROUGE metrics. Especially when programs share one common token with method names, our approach improves the absolute ROUGE-1 score by 7.8\%.
\end{abstract}


\section{Introduction}
Method names are particularly important during program development and maintenance. Semantically descriptive method names help developers reason about program behavior and further improve programming efficiency since complex program can be decomposed into several simple problems. 
In contrast, inconsistent method names confuse developers on the method usage and even lead to misuses and defects, as a result the program is hard to understand and maintain \cite{b1}\cite{b2}. 

\begin{figure}[htbp]
	\centerline{\includegraphics[width=0.48\textwidth]{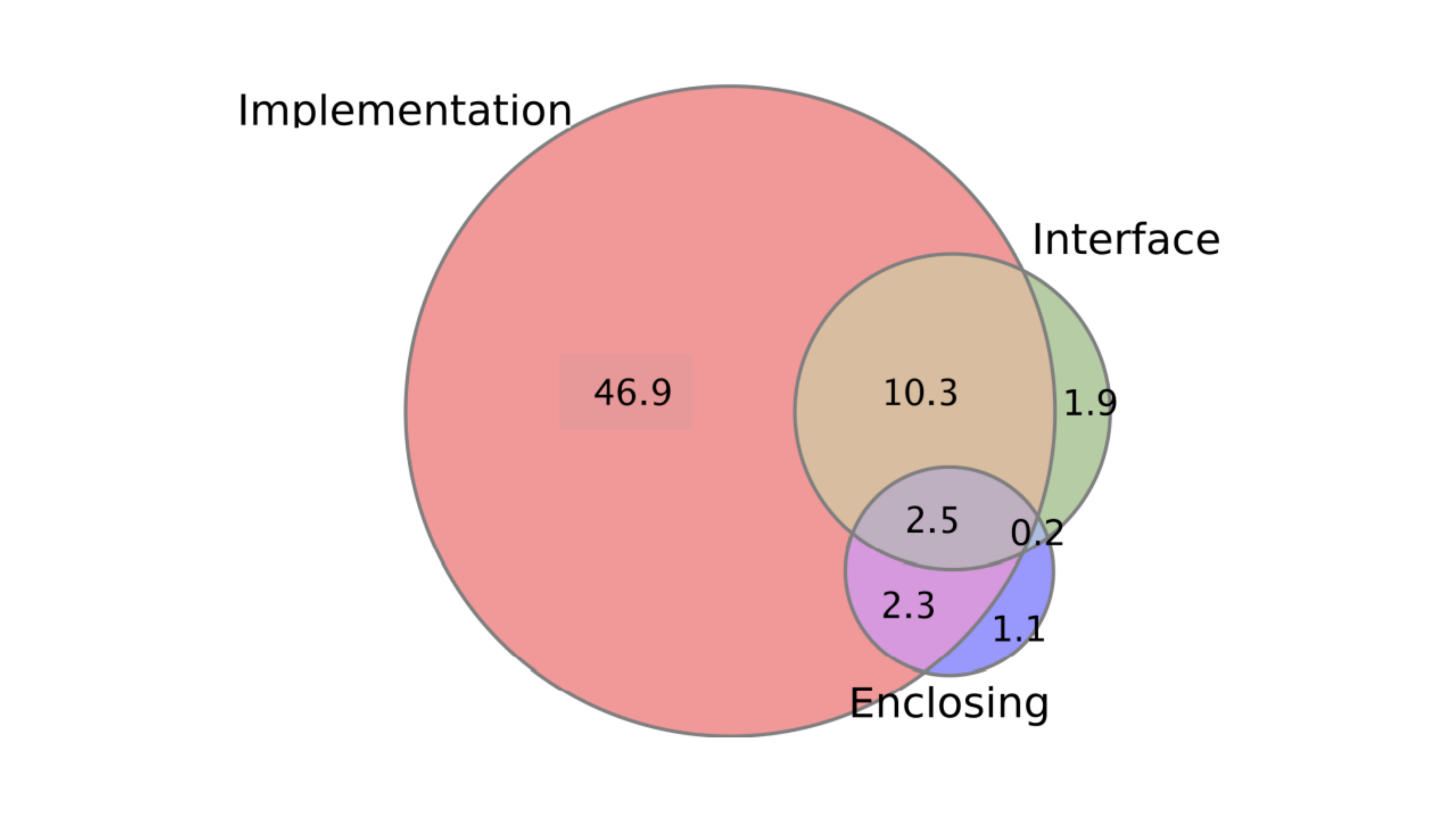}}
	\caption{ \% of tokens in method name found in contexts.}
	\label{fig}
\end{figure} 

Although development teams recognize the positive effect of meaningful method names, improper method names are common due to various reasons, such as inappropriate code cloning and conflicting naming conventions during collaboration among different developers \cite{b3}. Additionally, an empirical study with more than 13000 Java developers shows that renaming method is the most frequently refactoring operation, accounting for 74.8\% among all applied refactoring operation during program development \cite{b4}. Therefore, the automatic suggestion of descriptive method names will greatly improve the efficiency and quality of development.

To tackle the above issues, many studies have been investigated to suggest method names based on method implementation. Early heuristic or template-based approaches suggest method names according to several constructed rules and static features \cite{b5}\cite{b6}. However, the effectiveness of these early approaches depends on the constructed rules, which are not suitable for arbitrary program languages. Source code is unambiguous and highly structured, which can be reconstructed definitely according to its abstract syntax tree (AST). Therefore, many researchers explore the similarity of structure information contained in AST to automatically suggest method names \cite{b7,b8,b9}. Besides, the method naming task can be considered as a summarization problem of source code, which is much similar to the abstractive summarization problem in natural language processing (NLP) field. Thus, many researchers build their models based on the sequence-to-sequence \cite{b10} framework to generate summarized method names \cite{b11,b12,b13,b14}.

Recently, Nguyen et al.\cite{b11} conduct an empirical experiment on a large dataset used in the prior work \cite{b15} consisting of more than 17M methods in +14K highly-rated GitHub projects to explore the ratio of common tokens shared between a method name and the context. As seen in Fig.1, high proportions of the tokens in method name can be found in the three contexts of the corresponding method including its implementation, the interface and the enclosing class name, accounting for 62.0\%, 14.9\% and 6.1\% respectively. Thus, about 2 out of 3 tokens of a method name can be found in the three contexts, particularly in the method implementation. This observation indicates that the tokens composing the method name are closely related to the entity name within its method implementation. Based on this key finding, we aim to exploit these common shared token information between method name and method implementation to automatically generate method names.

In this paper, we decompose the method naming task into two related tasks, including keywords extraction task and method name generation task. Then we propose a keywords extractor and a keywords guided method name generation model to solve the two subtasks respectively. Specifically, 1) to sufficiently exploit the common shared token information, we design a graph neural network based supervised keywords extractor. The input of keywords extractor is a code graph constructed by the token sequence of code snippet and tree nodes of AST, and the output of keywords extractor is a set of keywords contained in source code. 2) To ensure the effectiveness of keywords guided method name generation, we adopt two strategies in encoder and decoder. 
We take the same GNN-based structure applied in keywords extractor as the encoder of method name generation model. The intuitive idea behind the GNN-based encoder is the powerful ability of graph neural network to capture syntactic and semantic information of source code simultaneously \cite{b14}.
The first strategy is the dual selective gate in encoder, which can control the information flow by exploring the relevance of input tokens with keywords. The second strategy is the dual attention mechanism employed in decoder, which can incorporate the semantics of input token sequence and keywords simultaneously, thus decoder can focus more on the keywords. 
Finally, we conduct extensive experiments on an open source dataset for method naming task. The experiment results show that our proposed model outperforms other state-of-the-art baselines significantly on the ROUGE metrics.

This paper makes the following contributions:
\begin{itemize}
\item To the best of our knowledge, we are the first to propose using keywords extraction task to improve the method naming task, and we design a supervised keywords extractor to address the keywords extraction problem.
\item We extend the graph-to-sequence model with a dual selective gate and a dual attention mechanism to effectively incorporate the key information contained in keywords.
\item We carry out extensive experiments on an existing dataset. Experiment results show that our model can improve the method naming task significantly compared to state-of-the-art models.
\end{itemize}

The remainder of this paper is organized as follows: Section II discusses the related work. Section III introduces the details of our proposed approach. Section IV shows the experiment settings and the evaluation results. Section V discusses the findings of generated method names. Finally, Section VI concludes the paper.

\section{Related Work}
Proper method names should summarize the key functionality of method as accurately and concisely as possible. While code comment generation task needs to generate meaningful and descriptive summarization of method, which is much similar to the method naming task. Therefore, in this section, we will discuss some of the related work from two aspects: method name generation and code comment generation.

\subsection{Method Name Generation}
Høst and Østvold \cite{b5} propose a heuristic based approach to check name consistency and suggest appropriate method names. They construct several explicit rules based on the prior knowledge of naming conventions. An inconsistency indicates a conflict between method name and the extracted rules. 
Liu et al. \cite{b17} design a deep learning based model to check the name consistency, and suggest method names based on the idea of information retrieval (IR), in such a way that two methods with similar implementations should have similar names. Allamanis et al. \cite{b18} introduce a machine learning based language model, which can explore the semantic similarity of method names by embedding them into a high-dimensional continuous space. 

Source code is unambiguous and highly structured, which can be reconstructed definitely according to given AST. Alon et al. \cite{b7} suggest method names via a general path-based approach which samples different paths in corresponding AST. Two methods that share similar bag-of-paths structure are close to each other. Later, Alon et al. \cite{b8} propose \emph{Code2vec}, a neural network model to aggregate the semantic information of AST paths, and produce a code vector using attention mechanism. Then the code vector can be used to generate method names. However, \emph{Code2vec} cannot represent unseen paths in training dataset. Alon et al. \cite{b9} further develop \emph{Code2seq} to represent paths node-by-node so that each unique path can be decomposed to a collection of nodes. Additionally, \emph{Code2seq} can generate neologisms which have not appeared in training corpus by splitting full tokens into subtokens. Jiang et al. \cite{b6} carry out extensive experiments to evaluate the effectiveness of \emph{Code2vec}, the evaluation results reveal that \emph{Code2vec} rarely works with more rigorous settings. They also propose a heuristic based approach to recommend method names.

With the powerful representation ability of neural network, many deep learning based models are devised to generate method names. Allamanis et al. \cite{b12} consider method names as the extreme summarization of source code, they propose a convolutional attention neural network that employs convolution on the code tokens to capture the structure information of source code. Nguyen et al. \cite{b11} adopt a sequence-to-sequence model with attention to predict method names, and the input tokens are collected from the three contexts including its body, the interface and the enclosing class name. 
Xu et al. \cite{b13} tackle method naming task by employing a hierarchical neural attention network to represent method implementation with the key idea that tokens form basic blocks and basic blocks form a code snippet. Fernandes et al. \cite{b14} propose a model for extending sequence encoders with a graph neural network. GNN-based model can mitigate the long-term dependencies in source code by specific edges. Besides, to resolve the out of vocabulary problem, they also incorporate pointer network \cite{b19} into their sequence decoder. 

\subsection{Code Comment Generation}
Code comment generation task is much similar to the abstractive sentence summarization of NLP. Thus, many strong models can be migrated to the code summarization task. Iyer et al. \cite{b20} establish the first Long Short Term Memory (LSTM) networks with attention to produce sentences that describe code snippets and SQL queries. Ahmad et al. \cite{b21} explore a transformer based model with relative position encoding to summarize source code. Transformer model has shown to be effective in capturing long-term dependencies \cite{b22}. Wei et al. \cite{b23} propose a dual training model to train the code summarization task and code generation task simultaneously to exploit the intuitive correlation between two tasks.

Above approaches just view source code as plain text, thus the structure information contained in AST are omitted. To address this problem, many studies have been investigated to incorporate the structure information into their models. Hu et al. \cite{b24} propose a structure based traversal method to traverse the AST. In such way, the structure information can be kept since the original AST can be unambiguously reconstructed from the traversal sequence. Wan et al. \cite{b25} apply an independent tree-based LSTM component to model the AST and solve the exposure bias issue by reinforcement learning. LeClair et al. \cite{b26} process code token sequence and code structure of AST as separate inputs, in this way the model can learn code structure independent of the token sequence. Furthermore, LeClair et al. \cite{b27} replace the gated recurrent unit with a graph neural network based component to model the AST. The experiment results show that graph neural network indeed improve the performance of the code summarization task.

The above approaches of method name generation and code comment generation have achieved remarkable performance. However, these deep learning models just implicitly learn the relations of important tokens in the source code with method name, while our keywords guided model can explicitly focus more on the valuable input.

\section{Method}
\begin{figure*}[htbp]
	\centerline{\includegraphics[width=1.0\textwidth]{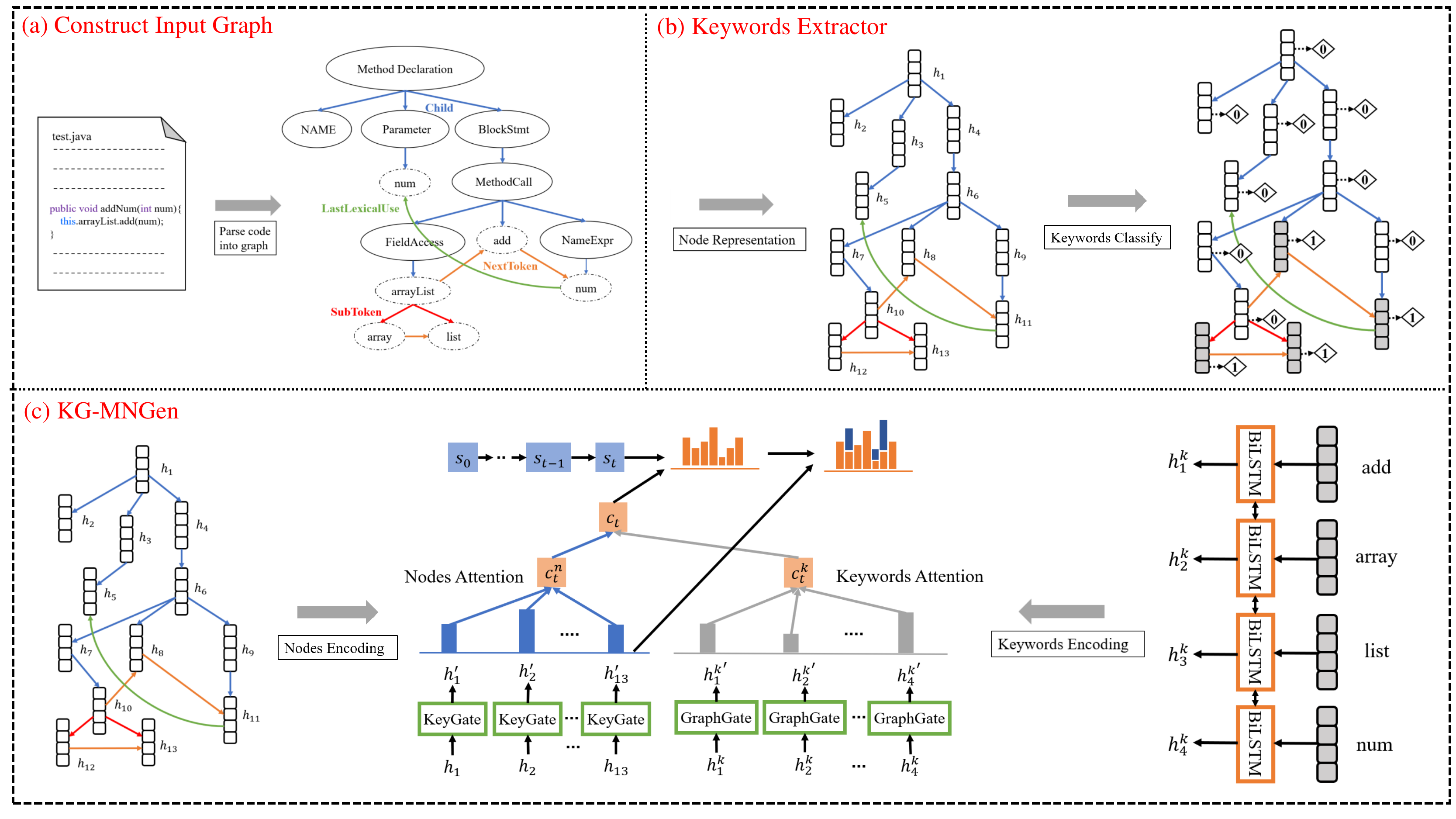}}
	\caption{ The architecture of our approach.}
	\label{fig}
\end{figure*}
\subsection{Overview}
The overall architecture of our keywords guided method name generation approach is depicted in Fig.2. We decompose the method naming task into two related tasks, and we propose a keywords extractor and a keywords guided method name generator to solve the two tasks respectively.

The motivation of our approach is the high proportions of common shared tokens between method name and method implementation, which can provide necessary clues for the key point of code snippet. 
For training, these keywords can be easily acquired by labeling the tokens appearing both in the input code token sequence and reference method name. For testing, we cannot directly obtain the reference method name to guide the labeling of keywords. Thus we propose a keywords extraction task, and we solve this task by training an independent keywords extractor in supervised way. Specifically, we apply a graph neural network based component to learn the representation of each token in the input graph, and a sigmoid layer over the node representation to decide whether a token is a keyword or not. 

Our keywords guided method name generator is based on a graph-to-sequence framework, which employs a same graph component as the keywords extractor to encode the representations of input tokens. 
Additionally, a standard LSTM based component is developed to learn the representation of each keyword. We introduce a dual selective gate to explore the semantic relationship between input tokens and keywords before decoding. For decoding, the copy mechanism is integrated in our sequence decoder, and we apply a dual attention mechanism to combine the semantics of input code sequence and keywords, while traditional attention mechanism only works on the input token sequence.

\subsection{Construction of Input Data}
The input of our keywords extractor is a graph constructed by source code tokens and corresponding AST. We label the overlapping tokens between the code tokens and the reference method name as 1, and other tokens as 0. The target of keywords extractor is essentially to classify each token into 1 or 0. Then we obtain the keywords by removing all the duplicated tokens predicted as 1. The keywords paired with the input graph of keywords extractor are fed into the method name generation model. We briefly introduce the details about the construction of input graph here.

An example for constructing the input graph is presented in Fig.2(a), following the works of Fernandes et al. \cite{b14}, we split all code tokens into subtokens according to the naming conventions of Java. For each code snippet, the node set is formed by all the tokens, subtokens and nodes of AST, with the original method name masked by a special symbol. Besides, we utilize several meaningful edges to connect each node based on the works of Fernandes et al. \cite{b14}. Firstly, we connect each node using \emph{NextToken} edge according to the order of the original source code sequence. Then the \emph{SubToken} edge is added to connect each token with its subtokens. To leverage the structure information, we also connect the nodes of the parse tree with its corresponding leaves by the \emph{Child} edge. Finally, we add \emph{LastLexicalUse} edge to connect identifiers to their most recent use in the source code.

\subsection{Graph Based Encoder}
Unlike natural language, source code contains complicated and easy-to-acquire structure information, which can be definitively represented by AST. Traditional sequence encoder views source code as plain text, and the rich structure information is neglected. While graph neural network can be directly applied to the input graph, thus the syntactic and semantic information of source code can be captured sufficiently.

Our graph encoder is based on the graph component proposed by Fernandes et al. \cite{b14}, which is developed according to the Gated Graph Neural Network (GGNN) \cite{b28}. A graph $G=(V,E,X)$ is consists of a node set $V$, a list of edge sets $E=(E_1,...,E_k)$, where $k$ is the number of edge types, and node embeddings $X$. Each node $u \in V$ corresponds to a node embedding $X_u \in R^{d_h}$, where $d_h$ is the dimension of the node embedding. The detailed procedure of message passing are as follows:

\begin{itemize}
	\item[1.] Each node $u$ needs to send the message to its neighbors. The message $m_u^{(t)}$ of each node $u$ is converted by an edge-type dependent function $f_k^{(t)}$ working on its current hidden state $h_u^{(t)}$. In our works, we use a simple linear layer to represent $f_k^{(t)}$ of edge type $k$ at timestep $t$. Note that we initialize the hidden state $h_u^{(0)}$ of each node $u$ by its associated node embedding $X_u$.
	\begin{equation}
	m_u^{(t)} = f_k^{(t)}(h_u^{(t)})
	\end{equation}
	
	\item[2.] Each node $u$ aggregates the received messages from its neighbors by element-wise summation. $N(u)$ means the neighbors of node $u$.
	\begin{equation}
	M_u^{(t)} = \sum_{v \in N(u)}m_v^{(t)}
	\end{equation}
	
	\item[3.] Each node $u$ updates its current state based on the aggregated messages $M_u^{(t)}$ , the update function is the gated recurrent unit (GRU) \cite{b29}.
	\begin{equation}
	h_u^{(t+1)} = GRU(M_u^{(t)}, h_u^{(t)})
	\end{equation}
\end{itemize}

The procedure of message passing is rolled out for timesteps $T$, and the hidden state of each node $u$ at the final timestep $h_u^{(T)}$ is used as node representation. In the following sections of this paper, we use $h_u$ to denote node representation to simplify the formula.

Besides, we obtain the global graph state $r_g$ by summing all the weighted node representations. The weight of each node is calculated based on the concatenation of the node representation $h_u$ and node embedding $X_u$, the details are given as follows:
\begin{equation}
r_g = \sum_{u \in V} \sigma (W_i[h_u:X_u]) \odot (W_jh_u)
\end{equation}
$W_i \in R^{2d_h \times 1}$, $W_j \in R^{d_h \times d_h}$ are two learnable parameter matrices, $\sigma(\cdot)$ is sigmoid function. Then we apply element-wise multiplication on the two outputs, and at last sum all the weighted node representations.

\subsection{Keywords Extractor}
After obtaining the node representations, the keywords extractor can determine whether a node is a keyword or not based on a simple dense layer with the sigmoid activation. We compute the cross entropy loss according to the output of sigmoid function $\hat{y}$ and ground truth labels $y$. The calculation details are given as follows:
\begin{equation}
\hat{y}^{(u)} = \sigma(W_eh_u+b_e)
\end{equation}
\begin{equation}
L_{key} = -\cfrac{1}{|V|} \sum_{u \in V} y^{(u)}log \hat{y}^{(u)}
\end{equation}
where $W_e$ is a projection matrix, $\sigma (\cdot )$ is sigmoid function, and $|V|$ is the size of node set. We minimize this loss function through Adam optimizer \cite{b30}.

\subsection{Dual Selective Gate}
We are inspired by the works of Li et al.\cite{b31} to encode the node representations and keywords in the dual selective way. The dual selective gate is expected to explore the relevance between the node representations and keywords. Therefore, the unnecessary information contained in node representations can be filtered out by the key selective gate since the keywords state contains more salient semantics. While the keywords representations can be adaptively adjusted by the graph selective gate since the graph state contains more global semantics.  

In our works, we use a standard bidirectional LSTM to encode the keyword representations. The BiLSTM reads the keyword embeddings and encode keywords forwardly and backwardly to generate hidden states: 
\begin{equation}
\overrightarrow{h^k_i} = LSTM(X_i, \overrightarrow{h^k_i}_{-1})
\end{equation}
\begin{equation}
\overleftarrow{h^k_i} = LSTM(X_i, \overleftarrow{h^k_i}_{+1})
\end{equation}
$X_i$ is the embedding of $i_{th}$ keyword, and the $i_{th}$ keyword representation $h^k_i$ is the concatenation of the forward and backward hidden states:
\begin{equation}
h^k_i = [\overrightarrow{h^k_i}:\overleftarrow{h^k_i}]
\end{equation}
Then we concatenate the representation of the first keyword $h^k_1$ and the last keyword $h^k_n$ to obtain the keywords state $r_k$:
\begin{equation}
r_k = [h^k_1:h^k_n]
\end{equation}
\textbf{dual selective gate} The encoded keywords state contains more salient semantics of input, which can guide the encoder to pay more attention on the important node representations. Meanwhile, the information contained in the keywords is incomplete, so more global semantic information contained in graph state is required to adaptively adjust the keyword representations. Therefore we apply a dual selective gate to explore the latent relevance between graph nodes and keywords. The calculation details are given as follows:
\begin{equation}
GraphGate_i = \sigma(W_kh^k_i + U_kr_g)
\end{equation}
\begin{equation}
KeyGate_u = \sigma(W_gh_u + U_gr_k)
\end{equation}
where $W_k, U_k, W_g, U_g \in R^{d_h \times d_h}$ are parameters, $GraphGate_i$ is the selective gate for $i_{th}$ keyword, and $KeyGate_u$ is the selective gate for node $u$. Note that the selective gate vector has the same dimension as the node representations as well as keyword representations. Then, we encode the node representations and keyword representations by computing the element-wise multiplication on the selective gate and original representations.
\begin{equation}
h^\prime_u = h_u \odot KeyGate_u
\end{equation}
\begin{equation}
h^{k^\prime}_i = h^k_i \odot GraphGate_i
\end{equation}

\subsection{Dual Attention Decoder}
Our decoder is based on the LSTM unit, and we use the graph state $r_g$ to initialize the decoder state. To sufficiently exploit the salient information contained in encoded keyword representations, we apply a dual attention mechanism to integrate the node information in input graph and keyword information. Besides, Out-of-vocabulary (OOV) tokens are ubiquitous in source code, thus we incorporate copy mechanism employed in pointer generator \cite{b16} to solve the OOV problem.

A general sequence decoder calculates hidden state $s_t$ according to the embedding of current input token $X_t$ and previous hidden state $s_{t-1}$ at timestep $t$.
\begin{equation}
s_t = LSTM(X_t, s_{t-1})
\end{equation}

Then we compute the attention scores of nodes in input graph based on the $s_t$ and node representations $h^\prime_u$, and obtain the context vector $c^n_t$ of nodes by weighted summing all the node representations.
\begin{equation}
e_u^t = W^\prime_a(tanh(W_a[s_t:h^\prime_u] + b_a))
\end{equation}
\begin{equation}
\alpha^t = softmax(e^t)
\end{equation}
\begin{equation}
c^n_t = \sum_{u \in V}\alpha_u^th^\prime_u
\end{equation}
where $W^\prime_a \in R^{d_h \times 1},W_a \in R^{2d_h \times d_h}$ are parameters, $b_a$ is bias. $\alpha^t$ is the attention score of current hidden state to each node representation. 

Similarly, we use hidden state $s_t$ and keyword representations $h^{k^\prime}_i$ to calculate the attention scores of keywords, and compute the context vector $c^k_t$ of keywords. It is noteworthy that the parameters of keywords attention mechanism are independent to the nodes attention mechanism.

Then we perform a linear transformation $U_c \in R^{2d_h \times d_h}$ on the concatenation of the nodes context vector $c^n_t$ and the keywords context vector $c^k_t$ to obtain the final context vector of current decoding step.
\begin{equation}
c_t = U_c[c^n_t:c^k_t]
\end{equation}

The output probability distribution of current timestep $t$ can be calculated based on the hidden state $s_t$ and context vector $c_t$.
\begin{equation}
P_{vocab} = softmax(W^\prime_v(W_v[s_t:c_t]+b_v)+b^\prime_v)
\end{equation}
where $W^\prime_v \in R^{d_h \times |vocab|}$, $W_v \in R^{2d_h \times d_h}$, $b^\prime_v \in R^{|vocab|}$, $b_v \in R^{d_h}$ are learnable parameters, $|vocab|$ denotes the vocabulary size. $P_{vocab}$ is a probability distribution of the output, and the output dimension corresponds to vocabulary size which informs decoder which token in the vocabulary should be used as the output at current decoding step.

Since the OOV tokens are common in source code, $P_{vocab}$ cannot contain the probability of OOV tokens. We refer to pointer generator \cite{b16} to copy the OOV tokens based on the attention distribution of input tokens. The pointer generator uses a soft switch to decide whether to copy a token from input token sequence or to output a token from vocabulary. The calculation details of soft switch are given as follows:
\begin{equation}
P_{gen}^t = \sigma (W_ss_t + W_cc_t + W_xX_t + b_{gen})
\end{equation}
where $W_s, W_c, W_x \in R^{d_h \times 1}, b_{gen} \in R^1$. At decoding step $t$, decoder reads hidden state $s_t$, context vector $c_t$ and embedding of the input token $X_t$ to produce a scaler
$P_{gen}^t \in [0,1]$. $P_{gen}^t$ denotes the probability for generating a token according to the vocabulary distribution $P_{vocab}$, and $1-P_{gen}^t$ denotes the probability for copying a token based on the attention distribution $P_{copy}$ where each attention score points to a token in source code. Thus the final output probability distribution is calculated as follows:
\begin{equation}
P_{copy}(w) = \sum_{u:w_u=w}\alpha_u^t
\end{equation}
\begin{equation}
P(w)=P_{gen}^tP_{vocab}(w)+(1 - P_{gen}^t)P_{copy}(w)
\end{equation}

At each decoding step, decoder outputs a token based on the $P(w)$, the loss function is the negative log likelihood of the generated token. Then the whole encoder-decoder model is trained in an end-to-end way, the overall loss for generated method names is:
\begin{equation}
loss = \cfrac{1}{T} \sum_{t=0}^{T} -log(P(w))
\end{equation}

\section{Experiments}
In this section, we first describe the dataset, evaluation metrics. Then we introduce several baselines and corresponding experiment settings. Finally, we report several research questions (RQs) and experiment results. 
\subsection{Dataset and Evaluation Metrics}
\noindent\textbf{Dataset.} We evaluate our approach on the Java-small dataset \cite{b9}, which contains 11 Java projects collected from GitHub. The Java-small dataset is widely used in method naming task, and the evaluation process can be divided into two ways:
\begin{itemize}
\item Train and test across projects, such as Alon et al. \cite{b9} and Fernandes et al.\cite{b14}.
\item Train and test across files, such as Allamanis et al.\cite{b12} and Xu et al. \cite{b13}.
\end{itemize}
We evaluate our approach across projects since the split-by-project choice prevents information leakage and other duplicates between different files from the same project. Therefore the
\begin{table}[htbp]
	\caption{Statistics For Java-small Dataset}
	\begin{center}
		\begin{tabular}{|c|c|c|c|c|c|}
			\hline
			\multirow{2}{*}{\textbf{Dataset}} & \multicolumn{3}{c|}{\textbf{Number of Examples}}                           \\ \cline{2-4} 
			& \textit{\textbf{Train}} & \textit{\textbf{Test}} & \textit{\textbf{Valid}} \\ \hline
			Java-small                        & 691489                  & 57016                  & 23837                   \\ \hline
		\end{tabular}
		\label{tab1}
	\end{center}
\end{table}
generalization ability of the model evaluated across projects is better than the model verified across files. To ensure the fairness of comparative experiments, we reuse the train-validation-test splits originally divided, 9 for training, 1 for validation and 1 for testing. More statistics about the Java-small dataset are presented in Table \uppercase\expandafter{\romannumeral1}.

\noindent\textbf{Evaluation Metrics.} We use ROUGE-N (N=1,2) and ROUGE-L \cite{b32} to evaluate the performance of our approach. Specifically, ROUGE-N counts the occurrence number of n-gram (n=1,2) in generated method name and reference method name, then outputs the F1-score. ROUGE-L is much similar to ROUGE-N, but it uses longest common subsequences to calculate the F1-score, instead of n-gram. In particular, ROUGE-1 is equivalent to F1 score which is used by Allamanis et al. \cite{b18} and Alon et al. \cite{b9}. ROUGE metrics are widely used to evaluate code summarization task.

\subsection{Baselines}
We compare our approach with following baselines:
\begin{itemize}
\item \textbf{ConvAttention.}  Allamanis et al. \cite{b12} use a convolutional attention neural network to summarize the code snippet.
\item \textbf{Code2seq.} Alon et al.\cite{b9} aggregate several sampled paths of AST and use attention mechanism to select relevant paths to suggest method names.
\item \textbf{BiLSTM+Copy.} Fernandes et al.\cite{b14} generate method names by incorporating sequence-to-sequence model with the copy mechanism.
\item \textbf{GNN+Copy.} Fernandes et al.\cite{b14} propose a graph-to-sequence framework to model the syntactic and semantic information of source code.
\item \textbf{BiLSTM+GNN+Copy.} Fernandes et al.\cite{b14} extend sequence-to-sequence model with a graph component to capture the structure information, which achieves state-of-the-art results on the method naming task.
\end{itemize}

\subsection{Experiment Settings}
In our experiment, we use 256-dimensional hidden state to represent input tokens and keywords. The gated graph neural network is rolled out for 4 timesteps. We remove all the tokens whose frequency of occurrence is less than five to build the vocabulary. When training, the correct keywords can be easily acquired, but the mismatching problem between training and testing leads to poor performance. Therefore we use the keywords predicted by the keywords extractor as the input to train the whole model. We initialize the Adam optimizer with the learning rate 0.0005 and decay the learning rate
\begin{table}[htbp]
	\caption{Evaluation Results on The Test Dataset}
	\begin{center}
		\begin{tabular}{|c|c|c|c|}
			\hline
			\multirow{2}{*}{\textbf{Models}} & \multicolumn{3}{c|}{\textbf{ROUGE}}                                               \\ \cline{2-4} 
			& \textit{\textbf{ROUGE-1}} & \textit{\textbf{ROUGE-2}} & \textit{\textbf{ROUGE-L}} \\ \hline
			ConvAttention                    & 33.1                     & -                         & -                         \\ \hline
			Code2seq                         & 43.0                      & -                         & -                         \\ \hline
			BiLSTM+Copy                      & 42.5                      & 22.4                      & 45.6                      \\ \hline
			GNN+Copy                         & 50.5                      & 24.8                      & 48.9                      \\ \hline
			BiLSTM+GNN+Copy                  & 51.4                      & 25.0                      & 50.0                      \\ \hline
			KG-MNGen                         & \textbf{53.4}                      & \textbf{26.4}                      & \textbf{53.6}                      \\ \hline
		\end{tabular}
	\end{center}
\end{table}
by 0.95 per 3000 steps. More experiment settings about our approach can be found in publicly available code repository \footnote{https://github.com/css518/Keywords-Guided-Method-Name-Generation}.

To ensure the fairness of the comparative experiments, we keep the hyper parameters consistent with our approach and train the state-of-the-art model with their original implementations. The evaluation results are similar as their original paper reported, thus we reuse the experiment results presented in their paper.

\subsection{RQ1: The Effectiveness of Our Approach}

Our proposed approach is referred to \textbf{KG-MNGen} (\textbf{K}eywords \textbf{G}uided \textbf{M}ethod \textbf{N}ame \textbf{Gen}erator). We compare \textbf{KG-MNGen} with several strong baselines under the ROUGE-1, ROUGE-2 and ROUGE-L evaluation metrics. The evaluation results are presented in Table \uppercase\expandafter{\romannumeral2}. We have run the \textbf{KG-MNGen} for many times and the fluctuation of experiment results is slight, around 0.3\%. Therefore the average results are reported. 

We can see from the table that \textbf{ConvAttention} obtains the lowest results among several baselines, which might due to the split-by-projects choice. \textbf{ConvAttention} is devised to predict method names across-files in its original paper. Additionally, all the graph based baselines achieve higher results than sequence based baselines. This observation indicates that effective modeling of structure information indeed boost the generation of accurate method names. Finally, \textbf{KG-MNGen} achieves the highest results on all the evaluation metrics. Compared with the state-of-the-art model \textbf{BiLSTM+GNN+Copy}, \textbf{KG-MNGen} improves absolute ROUGE-1, ROUGE-2 and ROUGE-L score about 2.0\%, 1.4\% and 3.6\% respectively. We argue that the reason is the definition of ROUGE-L, which uses longest common subsequences to calculate the evaluation score.  \textbf{KG-MNGen} can locate the key information effectively, thus \textbf{KG-MNGen} generates method names centered around the keywords, and the common subsequence is relatively longer.

In summary, \textbf{KG-MNGen} obtains the highest results among several strong baselines on all the evaluation metrics. The significant gains brought by \textbf{KG-MNGen} show that keywords guidance can indeed boost the effectiveness of method name generation.

\subsection{RQ2: The Effectiveness of Each Strategy}
To make full use of the positive signals brought by keywords, we adopt two strategies in the
\begin{table}[htbp]
	\caption{Evaluation Results of Each Strategy}
	\begin{center}
		\begin{tabular}{|c|c|c|c|}
			\hline
			\multirow{2}{*}{\textbf{Models}} & \multicolumn{3}{c|}{\textbf{ROUGE}}                                               \\ \cline{2-4} 
			& \textit{\textbf{ROUGE-1}} & \textit{\textbf{ROUGE-2}} & \textit{\textbf{ROUGE-L}} \\ \hline
			KG-MNGen                         & \textbf{53.44}            & \textbf{26.43}            & \textbf{53.58}            \\ \hline
			w/o KeyGate                      & 53.09                     & 26.17                     & 53.19                     \\ \hline
			w/o GraphGate                    & 53.12                     & 26.32                     & 53.24                     \\ \hline
			w/o DualGate                     & 52.98                     & 26.37                     & 53.12                     \\ \hline
			w/o DualAtten                    & 53.10                     & 25.79                     & 53.25                     \\ \hline
			w/o DualGate+DualAtten           & 51.82                     & 25.34                     & 51.63                     \\ \hline
		\end{tabular}
	\end{center}
\end{table}
encoder and decoder of \textbf{KG-MNGen}. One is the dual selective gate applied in encoder, which
consists of two gates, including key gate and graph gate. Another is the dual attention mechanism applied in decoder. Therefore, to sufficiently explore the different effects of each strategy, we compare \textbf{KG-MNGen} with several variants to understand the influence of dual attention mechanism and dual selective gate. The detailed evaluation results are reported in Table \uppercase\expandafter{\romannumeral3}. 

As we can see from Table \uppercase\expandafter{\romannumeral3}, firstly, the complete model \textbf{KG-MNGen} is superior to all the comparative variants under all the evaluation metrics, which indicates the effectiveness of the dual selective gate and dual attention mechanism. 

Secondly, compared to the \textbf{w/o DualGate+DualAtten}, both the dual selective gate equipped model \textbf{w/o DualAtten} and the dual attention mechanism equipped model \textbf{w/o DualGate} have made significant improvements under all the evaluation metrics. The experiment results prove that dual selective gate and dual attention mechanism can be applied separately, both the dual selective gate and dual attention mechanism are helpful to integrate the keywords signals.

Thirdly, we compare the influence of KeyGate and GraphGate respectively. The evaluation results of \textbf{w/o DualGate} are slightly lower than \textbf{w/o KeyGate} and \textbf{w/o GraphGate}, which proves the usefulness of each selective gate. And the performance improvements brought by KeyGate are slightly higher than GraphGate, which might due to the key information contained in the keywords, while GraphGate contains more comprehensive information. 

In summary, both the dual selective gate and dual attention mechanism are effective to incorporate the positive keywords guidance.

\subsection{RQ3: The Effects of Different Input Lengths on The Results}
We notice that the length of samples in the test dataset differ seriously, thus we compare
\begin{figure}[htbp]
	\centerline{\includegraphics[width=0.6\textwidth]{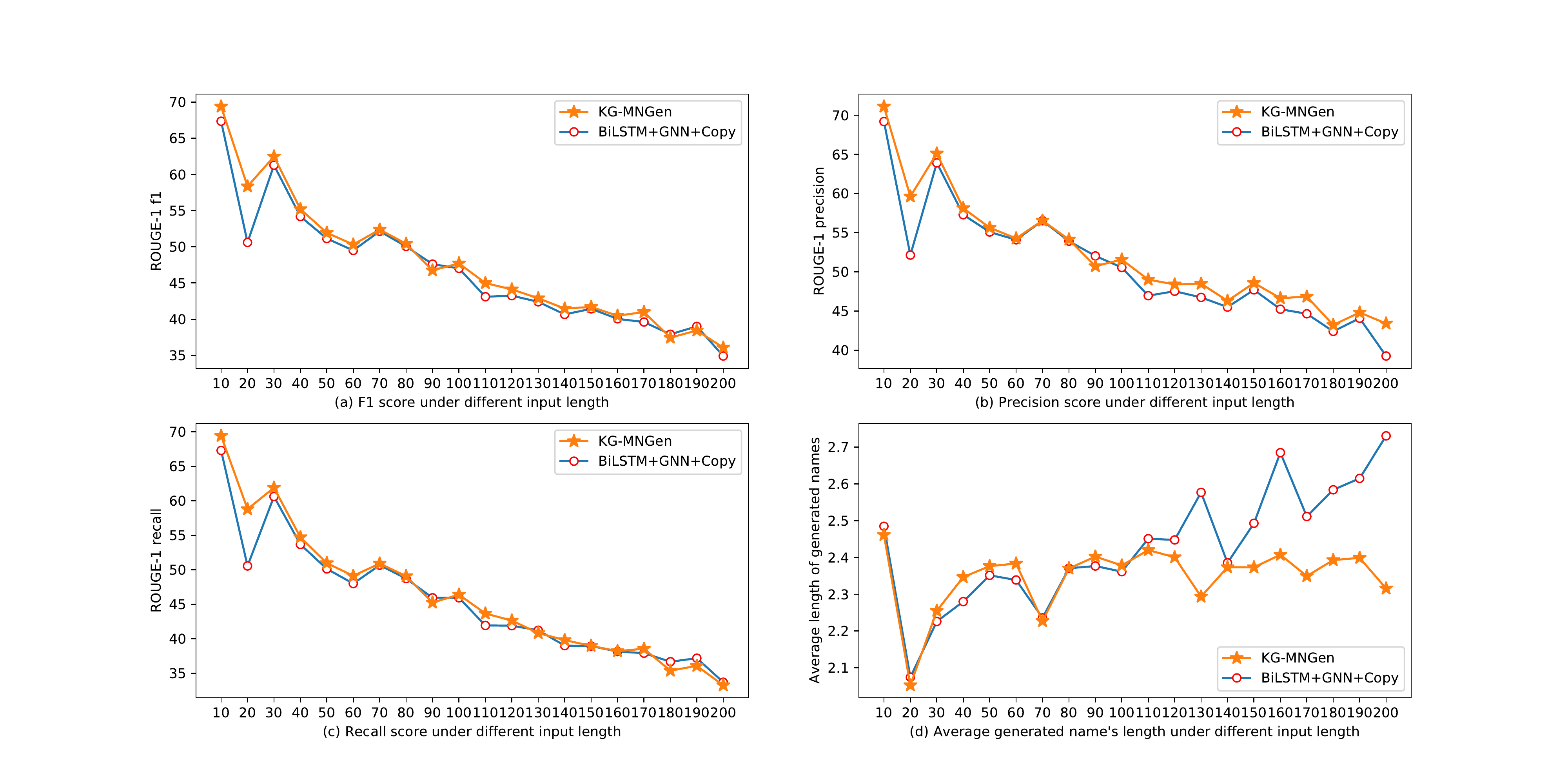}}
	\caption{ The f1, precision, recall of ROUGE-1 and average length of generated names under different input lengths.}
	\label{fig}
\end{figure}
\textbf{KG-MNGen} with the state-of-the-art model \textbf{BiLSTM+GNN+Copy} under different input length in terms of the f1, precision and recall of ROUGE-1 to explore the fine-grained effects of
different input lengths. The detailed results are presented in Fig.3. We use the number of the code tokens to denote the input length.

Firstly, all models show a natural descent as the input length increases. \textbf{KG-MNGen} outperforms \textbf{BiLSTM+GNN+Copy} almost across all input lengths, especially when input length between 20 and 30 tokens. Actually, input less than 30 tokens accounts for 47.3\% of the test dataset. Secondly, as we can see from Fig.3 (b) and Fig.3 (c), the improvements of precision score are higher than recall score, which might due to the key information contained in keywords, the decoder can generate more accurate tokens. Thirdly, we also count the number of generated tokens, which are reported in Fig.3 (d). \textbf{BiLSTM+GNN+Copy} is prone to generate longer method name as the input length increases, while \textbf{KG-MNGen} tends to generate stable length for method name. This observation can explain why two models obtain similar recall score. Higher recall means more tokens in reference method name can be found in generated method name, thus it becomes simple when more tokens are generated.

In summary, our approach \textbf{KG-MNGen} outperforms the strong baseline almost across all input lengths on the f1, precision and recall of ROUGE-1. The keywords can provide useful guidance under different input lengths.

\subsection{RQ4: Can Keywords Provide Effective Guidance to Other Baselines}

To explore the compatibility of keywords guidance methodology to other encoder-decoder frameworks, we equip two baselines with keywords signals, including \textbf{BiLSTM+Copy} and \textbf{BiLSTM+GNN+Copy}. The results are shown in Fig.4. All the evaluation metrics of two baselines have improved differently under the keywords guidance. In terms of \textbf{BiLSTM+Copy}, ROUGE-1 score improvements are roughly 3\%, while ROUGE-2 and ROUGE-L score improvements are slight. In terms of \textbf{BiLSTM+GNN+Copy}, the improvement of evaluation metrics are 1.1\%, 1.2\% and 2.6\% respectively. The experiment results are approximately linear related to the keywords.

\begin{figure}[htbp]
	\centerline{\includegraphics[width=0.5\textwidth]{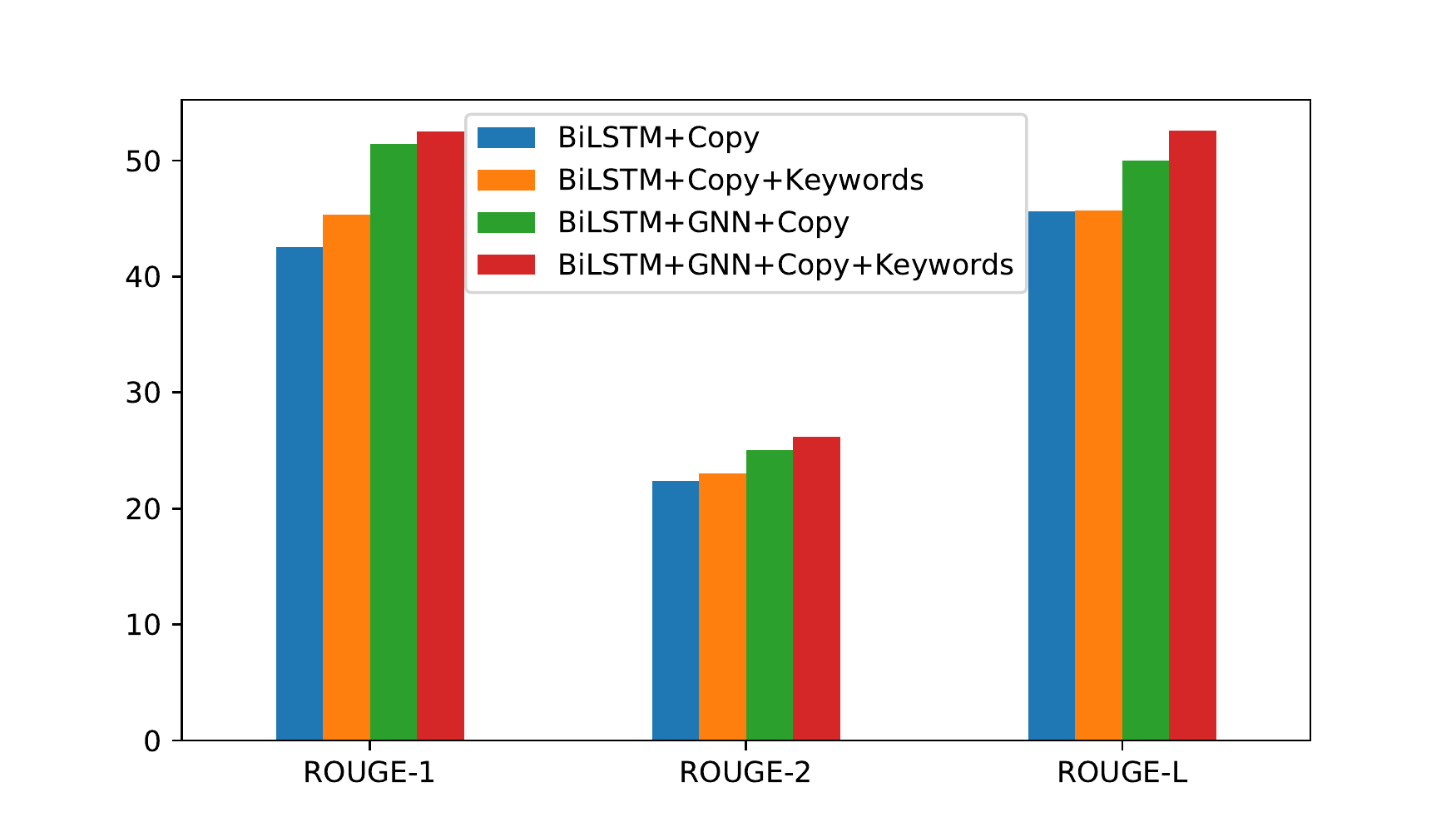}}
	\caption{ ROUGE score for two baselines equipped with keywords.}
	\label{fig}
\end{figure}

In summary, the keywords can provide effective guidance to other encoder-decoder framework, instead of just suitable for our proposed model.

\subsection{RQ5: The Effectiveness of Different Keywords}
Obviously, the performance of our approach is closely related to the quality of extracted keywords. Therefore, we extract keywords of method in different ways to explore the influence of the quality of keywords. The keywords selection strategies are as follows:
\begin{itemize}
	\item \textbf{Random.}  We randomly select 4 tokens from code token sequence as keywords. Note that we define several regular expressions to filter some inappropriate tokens, such as symbols and numbers, etc. 
	\item \textbf{TextRank\cite{b33}.} We select top 4 keywords according to the importance score computed by TextRank of each token.
	\item \textbf{TF-IDF.} We use TF-IDF to calculate the importance score of each token, instead of Textrank.
	\item \textbf{Keywords Extractor.} We train a GNN based keywords extactor in supervised way to extract the keywords of input.
	\item \textbf{Reference.} We directly use the reference method name of test dataset to label the common shared tokens as keywords, which is impossible in practice.
\end{itemize}

\begin{table}[]
	\caption{Evaluation Results of Different Keywords Selection Strategies}
	\begin{center}
		\begin{tabular}{|c|c|c|c|c|c|c|}
			\hline
			\multirow{2}{*}{\textbf{Keywords}} & \multicolumn{3}{c|}{\textbf{\begin{tabular}[c]{@{}c@{}}Keywords against\\ Ground Truth\end{tabular}}} & \multicolumn{3}{c|}{\textbf{\begin{tabular}[c]{@{}c@{}}Predictions against\\ Ground Truth\end{tabular}}} \\ \cline{2-7} 
			& \textit{\textbf{R-1}}           & \textit{\textbf{R-2}}          & \textit{\textbf{R-L}}          & \textit{\textbf{R-1}}            & \textit{\textbf{R-2}}           & \textit{\textbf{R-L}}           \\ \hline
			Random                             & 13.13                           & 0.70                           & 13.36                          &  52.59                                & 25.66                               & 52.72                               \\ \hline
			TextRank                           & 20.54                           & 1.13                           & 18.64                          & 52.70                                & 26.22                                & 52.85                                \\ \hline
			TF-IDF                             & 22.31                           & 1.69                           & 19.51                          & 52.77                                 & 26.17                                & 53.01                                \\ \hline
			Keywords Extractor                 & 36.57                           & 7.10                           & 32.15                          & 53.44                                 & 26.43                                & 53.58                                \\ \hline
			Reference                          & 67.87                           & 14.30                          & 54.54                          & 67.33                                 & 39.27                                & 67.60                                \\ \hline
		\end{tabular}
	\end{center}
\end{table}

The evaluation results are reported in Table \uppercase\expandafter{\romannumeral4}. We use the ROUGE score of extracted keywords against ground truth method name to indicate the quality of keywords. The experiment results are consistent with our intuition, random strategy obtains the worst results, TF-IDF strategy is slightly better than TextRank strategy, while keywords extractor strategy achieves a significant performance improvement compared to the unsupervised TF-IDF and TextRank strategy. However, there is still a very large performance gap relative to the reference strategy, which can be regarded as the upper bound of extracted keywords. 
Additionally, it is normal that ROUGE-2 scores of all experiments are extremely poor since the extracted keywords are independent of each other, thus they cannot form a highly readable method name. 

Then we input \textbf{KG-MNGen} with above different keywords to explore the correlation between the quality of keywords and model performance. The evaluation results are recorded in Table \uppercase\expandafter{\romannumeral4} \textbf{Predictions against Ground Truth}. Firstly, it can be concluded from the experiment results that the performance of the model is roughly proportional to the quality of the keywords, which also reveals the effectiveness of our keywords guided method name generation approach. Secondly, it is a little surprise to us that random strategy achieves slightly worse experiment results with TextRank. This is contrary to our assumptions since we originally think that random keywords might provide misleading guidance for the model and reduce the performance of the model. Thirdly, ROUGE-2 score of all experiments obtain a significant improvement, which due to the sequence decoding of language model. Higher ROUGE-2 score means highly readable and fluent method name. Lastly, reference strategy achieves promising results, far superior to other keywords selection strategy. However, reference strategy cannot be applied in actual since the reference method name of test dataset are unavailable. Nevertheless, we can optimize our keyword extractor to generate more accurate keywords. Note that our keywords extractor is relatively simple, only using graph neural network for representation. Thus there are still many aspects of keywords extractor can be enhanced, as a result more accurate keywords lead to better performance.

In summary, firstly, the keywords guided method name generation approach is highly related to the quality of keywords. Secondly, inspired by the high performance brought by reference strategy, the method name generation task can be improved significantly by optimizing the keywords extraction task. Additionally, compared to the method name generation task, the keywords extraction task is easier to optimize. But optimizing the keywords extraction task has a definite upper bound on performance.
\begin{table}[]
	\caption{ROUGE-1 Score of Different Number of Shared Tokens Between Our Approach and Baseline}
	\begin{center}
		\begin{tabular}{|c|c|c|c|c|c|}
			\hline
			\multirow{2}{*}{Results} & \multicolumn{5}{c|}{Number of Shared Tokens}              \\ \cline{2-6} 
			& =0    & =1             & =2    & =3    & \textgreater{}=4 \\ \hline
			Data Size                & 10860 & 18669          & 13644 & 8273  & 5557             \\ \hline
			BiLSTM+GNN+Copy          & 18.38 & 54.05          & 63.09 & 64.75 & 58.44            \\ \hline
			KG-MNGen                 & 18.77 & 60.51          & 63.44 & 64.61 & 58.68            \\ \hline
			Gains                    & +0.39 & \textbf{+6.46} & +0.35 & -0.14 & +0.24            \\ \hline
		\end{tabular}
	\end{center}
\end{table}

\begin{table}[]
	\caption{ROUGE-1 Score of Different Code Cases Between Our Approach and Baseline}
	\begin{center}
		\begin{tabular}{|c|c|c|}
			\hline
			Results         & Template Methods & Other Methods  \\ \hline
			Data Size       & 4484             & 14185          \\ \hline
			BiLSTM+GNN+Copy & 69.11            & 49.29          \\ \hline
			KG-MNGen        & 71.20            & 57.13          \\ \hline
			Gains           & +2.09            & \textbf{+7.84} \\ \hline
		\end{tabular}
	\end{center}
\end{table}

\section{Discussion}
\subsection{The Difference of Generated Method Names}
In this subsection, we present the discussion on the comparison of generated method names between \textbf{KG-MNGen} and the state-of-the-art baseline \textbf{BiLSTM+GNN+Copy} to qualitatively evaluate the superiority of our approach. 

\textbf{KG-MNGen} is based on the common shared token information between method body and reference method name. Therefore, we split the test dataset into five parts according to the number of shared tokens. Then we compute the ROUGE-1 score on each part of the test dataset. The detailed results are reported in Table \uppercase\expandafter{\romannumeral5}. 

Firstly, the performance of two approaches shows a stable increasing tendency when the number of shared tokens is less than 4. We can conclude that it is relatively easy to generate method names when more tokens are common shared since high proportions of key information are already explicitly contained in method body. Besides, it is normal that the ROUGE-1 score decreases when the number of shared tokens is greater than 4 since the average length of generated method names of two approaches is less than three tokens. Therefore, the ROUGE-1 score shows a natural descent as the lengths of method name increase.

Secondly, both approaches obtain poor performance when there are no shared tokens. It means that key information is not explicitly shown in method body. Therefore a descriptive method name requires the model to fully comprehend the program as well as developers. Unfortunately, it is much harder to achieve this. Besides, the performance of \textbf{\textbf{KG-MNGen}} is slightly higher than \textbf{BiLSTM+GNN+Copy} when the number of shared tokens is greater than 2. As aforementioned, it is relatively easy to understand the program when more common tokens are shared. Therefore, \textbf{BiLSTM+GNN+Copy} can achieve similar results as our approach in relatively simple cases.

\begin{figure*}[htbp]
	\centerline{\includegraphics[width=0.99\textwidth]{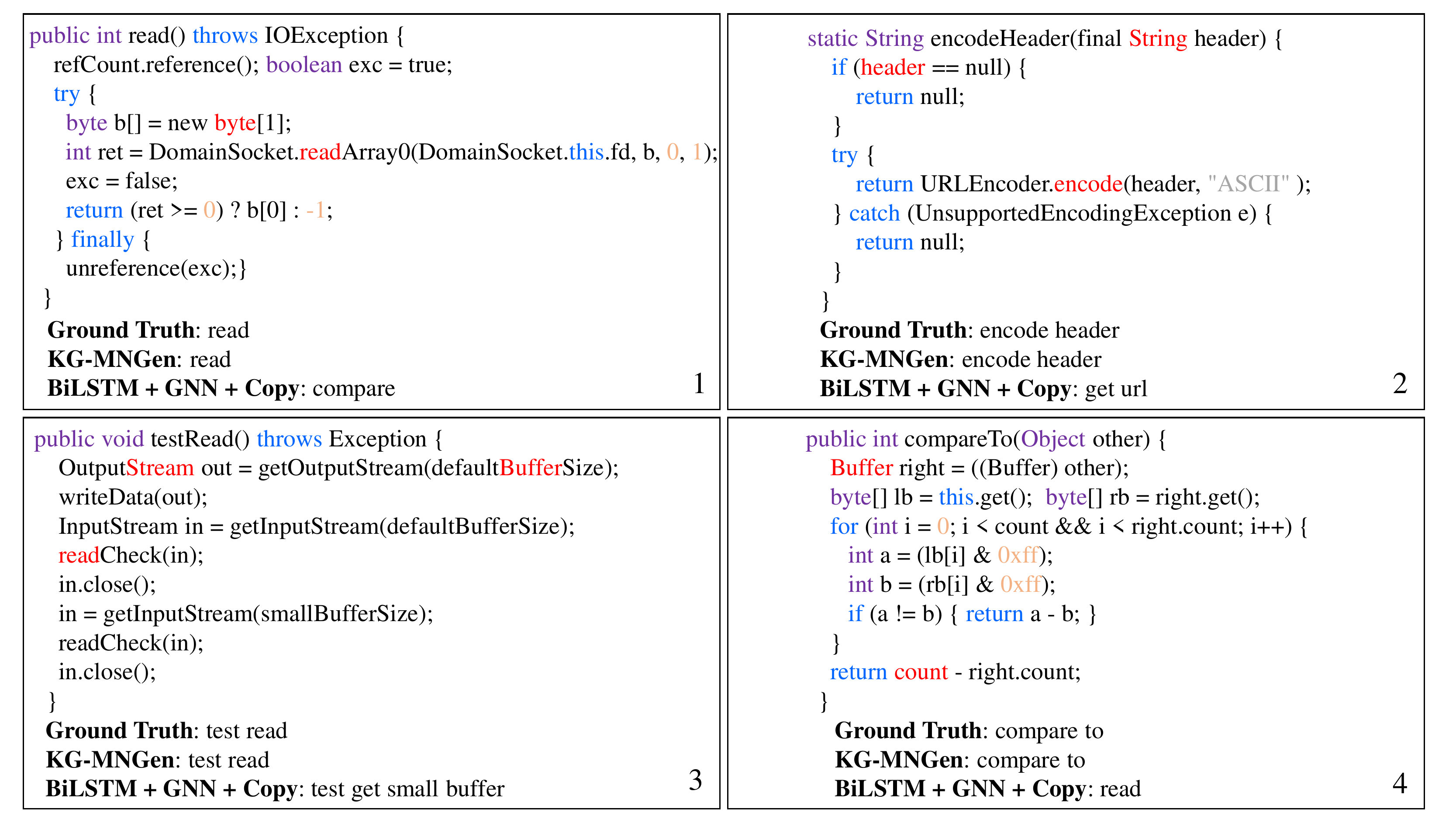}}
	\caption{ Some examples in test dataset}
	\label{fig}
\end{figure*}
Thirdly, it is hard to generate method names when just one common token is shared. A little key information is explicitly contained in method body. Thus the model needs to comprehend the program based on the key information. As we can see from Table \uppercase\expandafter{\romannumeral5}, \textbf{BiLSTM+GNN+Copy} obtains poor results and has a wide margin compared with other experiments of \textbf{BiLSTM+GNN+Copy}, about 9.04\% and 10.7\% lower than the experiments of 2 and 3 shared tokens. While \textbf{KG-MNGen} bests \textbf{BiLSTM+GNN+Copy} by significant margins of 6.5\%. This finding indicates that \textbf{KG-MNGen} is much superior than \textbf{BiLSTM+GNN+Copy} in hard cases. Additionally, the size of each part of the test dataset shows that condition with only one shared token is more common. Therefore, we can conclude that our approach can generate more accurate method names in hard cases than \textbf{BiLSTM+GNN+Copy}.

Finally, we perform a more detailed comparison in the condition with only one shared token to evaluate how the 6.5\% improvements are presented in generated method names. As we all know, there are some template methods in Java program, such as \emph{setXxx}, \emph{getXxx}, \emph{toString}, \emph{equals} and \emph{hashCode},etc. These methods are easily to predict method names since they share similar structures. So we further verify our approach after removing above simple methods. The evaluation results are reported in Table \uppercase\expandafter{\romannumeral6}. Our approach gains an absolute ROUGE-1 score about 7.8\% over the baseline after removing above listed template methods. The wide margin indicates that our approach is more effective with rigorous settings compared with \textbf{BiLSTM+GNN+Copy}.

To sum up, in terms of the generated method names between \textbf{KG-MNGen} and \textbf{BiLSTM+GNN+Copy}, \textbf{KG-MNGen} can generate more accurate method names in different code cases. Especially in complex programs, \textbf{KG-MNGen} outperforms \textbf{BiLSTM+GNN+Copy} by significant margins of 7.8\%.

\subsection{Case Study}
We present some examples of generated method names by \textbf{KG-MNGen} and \textbf{BiLSTM+GNN+Copy} in Fig.5, and we mark the extracted keywords in red. Firstly, although some noise are contained in our keywords, \textbf{KG-MNGen} is able to focus more on the accurate keywords and generate correct method name. While \textbf{BiLSTM+GNN+Copy} is easily affected by the messy information in the source code, thus misunderstanding the key function of program, such as example 3. The function of example 3 is to test whether the program can read the input normally, but \textbf{BiLSTM+GNN+Copy} only focuses on the \emph{smallBuffer}. Secondly, none of tokens are shared in example 4. This case needs model to fully comprehend the program as well as developers. While \textbf{KG-MNGen} predicts the method name of example 4 accurately. Thirdly, example 1 denotes the condition with one shared token. \textbf{BiLSTM+GNN+Copy} might focus on the \emph{boolean} and \emph{ret$\textgreater$=0}, thus mistaking the key function of this program for \emph{compare}. While \textbf{KG-MNGen} predicts the accurate method name under the keywords guidance.

\section{Conclusion and Future Work}
In this paper, inspired by the high proportions of common shared tokens between method name and method implementation, we propose a keywords guided method name generation approach. Our approach contains two related tasks, including keywords extraction task and method name generation task. 
We devise a keywords extractor to optimize the keywords extraction task and a keywords guided method name generator \textbf{KG-MNGen} to predict method names. Comprehensive experiments compared to the current state-of-the-art baselines demonstrate that \textbf{KG-MNGen} can effectively generate more accurate method names, improving ROUGE metrics by 1.5\%-3.5\%. Besides, \textbf{KG-MNGen} outperforms the baseline by significant margins of 7.8\% when programs share one common token with method names. In the future, we plan to explore the methodology of incorporating more semantic information into keywords guided model.


\section*{Acknowledgment}
This work has been supported by the National Key R\&D Program of China under grant 2018YFB1003800,
National Natural Science Foundation of China (No.61772560), Natural Science Foundation of Hunan
Province (No. 2019JJ40388).

\balance

\end{document}